\documentclass[final,5p,times,twocolumn]{elsarticle}

\usepackage{amssymb} 
\usepackage[hyphens]{url}  
\usepackage{breakurl}
\usepackage{hyperref}
\usepackage{etoolbox} 
\usepackage{balance}
\usepackage{amsmath}
\usepackage{array}
\usepackage{fancyhdr}
\newcommand{\orcidlink}[1]{\textsuperscript{\href{https://orcid.org/#1}{\includegraphics[scale=0.2]{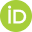}}}}

\journal{Pre-print}

\begin{document}

\begin{frontmatter}



\title{How LLMs are Shaping the Future of Virtual Reality}


\author[]{Süeda Özkaya\,\orcidlink{0009-0004-5954-5280}}
\author[]{Santiago Berrezueta-Guzman\,\orcidlink{0000-0001-5559-2056}}
\author[]{Stefan Wagner\,\orcidlink{0000-0002-5256-8429}}
\affiliation[]{organization={Technical University of Munich},
           city={Heilbronn},
            country={Germany}}

\begin{abstract}
The integration of Large Language Models (LLMs) into Virtual Reality (VR) games marks a paradigm shift in the design of immersive, adaptive, and intelligent digital experiences. This paper presents a comprehensive review of recent research at the intersection of LLMs and VR, examining how these models are transforming narrative generation, non-player character (NPC) interactions, accessibility, personalization, and game mastering. Drawing from an analysis of 62 peer-reviewed studies published between 2018 and 2025, we identify key application domains—ranging from emotionally intelligent NPCs and procedurally generated storytelling to AI-driven adaptive systems and inclusive gameplay interfaces. We also address the major challenges facing this convergence, including real-time performance constraints, memory limitations, ethical risks, and scalability barriers. Our findings highlight that while LLMs significantly enhance realism, creativity, and user engagement in VR environments, their effective deployment requires robust design strategies that integrate multimodal interaction, hybrid AI architectures, and ethical safeguards. The paper concludes by outlining future research directions in multimodal AI, affective computing, reinforcement learning, and open-source development, aiming to guide the responsible advancement of intelligent and inclusive VR systems.
\end{abstract}

\begin{keyword}
Accessibility\sep Affective Computing\sep AI Game Mastering\sep Ethical AI\sep Immersive Games\sep Large Language Models (LLMs)\sep  Memory Management\sep Multimodal Interaction\sep NPC Interaction\sep Personalized Gameplay\sep Procedural Storytelling\sep Real-Time Systems\sep Virtual Reality (VR).

\end{keyword}

\end{frontmatter}

\section{Introduction}\label{I}

The integration of Large Language Models (LLMs) into Virtual Reality (VR) games represents a transformative step in the evolution of interactive digital environments \cite{plupattanakit2024llmseduverse, wei2024collaborative}. LLMs are neural networks trained on extensive text data to produce language that resembles human speech. They have progressed quickly in their abilities and uses, evolving from text-generating tools to agents engaging in real-time dialogue, narrative design, and adaptive learning \cite{minaee2024llmsurvey, brown2020language}. In parallel, VR games have evolved from simulations to immersive worlds that leverage spatial computing, haptic feedback, and embodied interaction \cite{anthes2016state}. The convergence of these technologies offers unprecedented opportunities to enhance interactivity, narrative depth, emotional engagement, and accessibility in digital games \cite{alkhayat2024leveraging, tanksale2023web3d}.

The core focus of gaming and simulation development on immersive experiences has generated rising interest in artificial intelligence (AI) applications, especially LLMs for creating dynamic and human-like gameplay  \cite{sweetser2024llmsgames, sissler2024enhancing}. The exploration of LLMs as creative tools for virtual reality experiences has emerged because these models enable complex non-player character (NPC) dialogues and generate storylines and environments that adapt to player choices \cite{normoyle2024using, peng2024player}. The models' context-sensitive language capabilities enable personalized gameplay that becomes more accessible and inclusive, thus expanding educational and training possibilities and entertainment options \cite{voultsiou2025aiinclusive, bozkir2024embedding}.

This research paper explores the developing relationship between LLMs and VR games to determine their applications for improving core immersive gameplay elements. This research investigates the following key questions:  \\
\textbf{ - RQ1.} How do LLMs contribute to more emotionally intelligent and lifelike NPC interactions? \\
\textbf{ - RQ2.} In what ways can they support procedural storytelling and adaptive narratives? \\
\textbf{ - RQ3.} How do they affect personalization, accessibility, and user experience in immersive environments? \\
\textbf{ - RQ4.} What challenges and limitations -technical, ethical, and practical—must be addressed while achieving their full potential? \\
\textbf{ - RQ5.} How can future research leverage emerging trends in multimodal AI, reinforcement learning, affective computing, and open-source tools to build scalable, ethically responsible, and emotionally attuned VR systems?

To answer these questions, we conduct a comprehensive literature review that examines current research and systems across six application domains: (1) dynamic NPC interactions and emotional intelligence, (2) procedural storytelling and narrative generation, (3) intelligent game mastering and adaptive control systems, (4) personalized player experience, (5) accessibility, inclusivity, and usability, and (6) challenges and limitations including ethical concerns and deployment barriers. Additionally, we analyzed the current state of the art before defining the main opportunities and constraints forming the future direction of LLM-based VR gaming.

\begin{figure}[htbp]
    \centering
    \includegraphics[height=0.5\textheight]{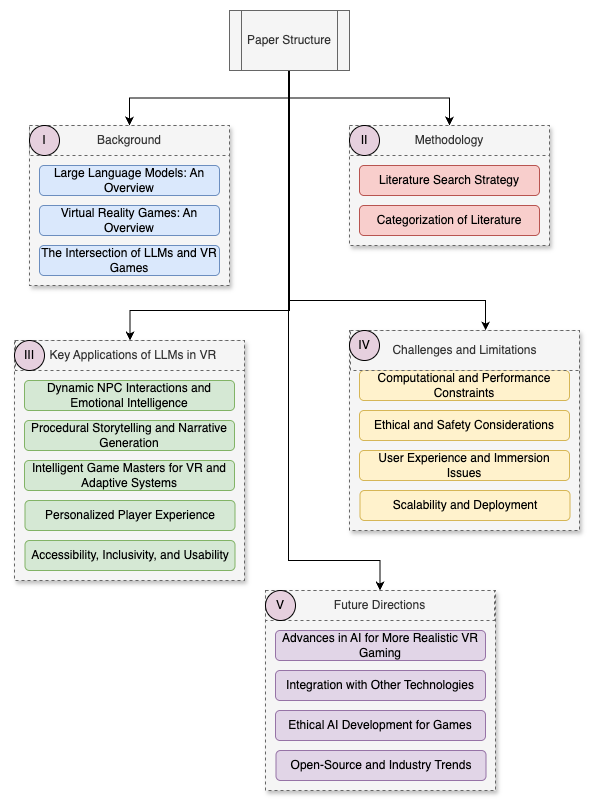}
    \caption{Overview of the paper structure and organization.}
    \label{fig:structure}
\end{figure}

This paper contributes to the expanding field of intelligent digital worlds by evaluating the combined impact of LLMs and VR on gameplay and critically reviewing their technological convergence. It aims to support researchers, developers, and designers seeking to build more engaging, equitable, and responsive VR environments through LLMs' responsible and creative use.
Figure~\ref{fig:structure} provides an overview of this review's organization and thematic flow.

\section{Background}

Understanding the integration of Large Language Models (LLMs) into Virtual Reality (VR) games requires a foundational overview of both technologies and their evolution.

\subsection{Large Language Models (LLMs)--Overview}

LLMs have experienced considerable advancement in recent years, moving from theoretical ideas to more advanced and scalable architectures like OpenAI's Generative Pre-trained Transformers (GPT) series, Meta's Large Language Model Meta AI (LLaMA), and Google's PaLM. The development of large-scale transformer-based architectures has facilitated significant progress in generating text, creating dialogue systems, and enabling procedural content generation \cite{minaee2024llmsurvey}. 

LLMs began with early autoregressive models, which focused on predicting the next word in a sentence based on the words that came before. At first, this approach was a small area of research in Natural Language Processing (NLP). However, it gained major attention after the release of GPT-2 in 2019, where it showed that transformer models trained on massive text datasets could generate high-quality, coherent language—and that their output could be shaped using carefully designed prompts \cite{radford2019language, gallotta2024large, du2024survey}. 

OpenAI’s GPT models have been leading in developing LLMs, driving major progress in natural language processing \cite{wang2024history}. GPT-3, with 175 billion parameters, impressed researchers with its ability to generate fluent and meaningful text, which led to its use in chatbots and interactive AI tools \cite{zong2022survey}. Later versions—GPT-3.5 and GPT-4—further improved these abilities by using Reinforcement Learning from Human Feedback (RLHF), which helped the models give more accurate and helpful responses \cite{brown2020language, du2024survey, hauser2024histllm}.

In addition to OpenAI’s work, other powerful transformer-based models have been developed. Bidirectional Encoder Representations from Transformers (BERT) introduced bidirectional training, which made it better at understanding context and improved tasks like text classification and natural language inference. LLaMA was trained only on publicly available data, making it more accessible for open-source use. While BERT focused on deep understanding of text, LLaMA aimed to offer smaller, more efficient models without losing performance compared to other top models \cite{devlin2019bert, touvron2023llama, singh2023exploring}.

As LLMs have grown in size and capability, new multimodal models like GPT-4V \cite{openai2024gpt4} and Large Language and Vision Assistant (LLaVA) \cite{liu2023visualinstructiontuning} have expanded their use beyond just text. These models can now understand images, generate speech, and support interactive storytelling. They are also being improved to use memory and computing power more efficiently while becoming more accurate and flexible.

\subsection{Virtual Reality Games--Overview}

Virtual Reality (VR) gaming has come a long way, moving from simple simulations to highly immersive and interactive experiences. This progress has been driven by hardware, software, and AI advances, especially in areas like NLP and content generation.

The beginnings of VR can be traced back to 1968, when Ivan Sutherland and his student Bob Sproull created the first computer-based VR system, known as the \textit{"Sword of Damocles"}. It used head tracking to display a basic 3D wireframe view that changed with the user’s perspective. While it wasn’t a game, it introduced key ideas, like perspective-based interaction, that still form the foundation of modern VR gaming \cite{adhikary2017origin, anthes2016state}.

The 3D Internet allows users to move around and interact with digital objects in space, rather than just clicking on flat screens. A well-known example is \textit{Second Life} \cite{zhu2007secondlife}, a virtual world where people can socialize and explore using avatars. It combines AI, 3D headsets, motion sensors, and even holographic displays to create more immersive experiences \cite{gfg3dinternet}.

After early VR experiments in the 1980s, the release of consumer-friendly devices like the Oculus Rift, HTC Vive, and PlayStation VR changed the gaming world. These headsets gave players access to realistic virtual worlds with features like motion tracking, haptic feedback (touch sensations), 3D visuals, and the use of photogrammetry \cite{berrezueta2025reality}.

As VR technology improved, games became more realistic, featuring better physics, stronger hardware, smarter characters, and stories that change based on player choices. Modern VR games now include AI-powered characters, voice interaction, and multiple communication methods, making the experience feel more natural and immersive, like talking to real people.

VR gaming hardware includes two main parts: output devices and input devices. Output devices, like head-mounted displays (HMDs), show 3D visuals and provide a wide field of view to make the experience realistic. These can be mobile (like Google Cardboard or Samsung GearVR) or wired (like Oculus Rift, HTC Vive, and PlayStation VR), and often include motion sensors and sometimes eye tracking for better performance. To simulate touch, devices like vests, gloves, and full-body suits provide haptic feedback, letting players feel things like force, wind, or temperature. All these tools work together using motion sensors and tracking systems to deliver accurate and responsive gameplay \cite{anthes2016state}.  

Despite its rapid development, the VR gaming sector still faces key challenges, including high production costs, limited content, motion sickness, and hardware accessibility. The immersive potential of VR is well recognized, but the expense of headsets and the need for high-performance computers continue to hinder widespread adoption \cite{anthes2016state}. Additionally, some users experience nausea or disorientation during extended play, often due to latency or mismatched sensory input \cite{laviola2000cybersickness}. Developing VR content also demands specialized tools and expertise, creating barriers for smaller studios. These factors highlight that technical and usability issues must be addressed for VR gaming to scale sustainably.

\subsection{The Intersection of LLMs and VR Games}

The convergence of LLMs and VR technologies represents a significant advancement in developing interactive digital environments, particularly in entertainment-based and serious games \cite{damianova2025seriousvr}. The convergence of these technologies enables developers to create innovative immersive experiences, dialogue systems, and educational simulations \cite{plupattanakit2024llmseduverse, wei2024collaborative}.

Initial applications of LLMs in VR have focused on improving user interaction and supporting developers. One of the earliest documented integrations involved helping novice developers create VR content. Tools like ChatGPT assist with code suggestions, debugging, and explaining concepts within development environments such as Unity, making the VR development process more accessible \cite{alkhayat2024leveraging}.

Beyond development support, LLMs are increasingly used in real-time VR environments to generate narrative content, enable dynamic interactions, and simulate intelligent NPCs. These conversational agents are especially valuable in educational and training applications, where they provide adaptive feedback and act as virtual tutors \cite{tanksale2023web3d, song2024learningverse}.

A key use of LLMs in VR is in AI-driven dialogue systems. Unlike scripted interactions, LLMs support unscripted, context-aware conversations that enhance immersion. For instance, the "LearningverseVR" platform \cite{song2024learningverse} uses generative AI to create NPCs with distinct personalities and backgrounds, enabling learners to engage in personalized, natural dialogue while exploring content at their own pace.

Various VR game genres have integrated LLM-based assistants to guide players through challenges using voice interaction \cite{rychert2024gptvr, tonini2024talk, lau2024scottishvr, bateni2024languagedriven}. These systems offer real-time problem-solving and adapt to player input, enhancing engagement through natural, human-like communication.

\begin{figure}[htbp]
    \centering
    \includegraphics[width=0.5\textwidth]{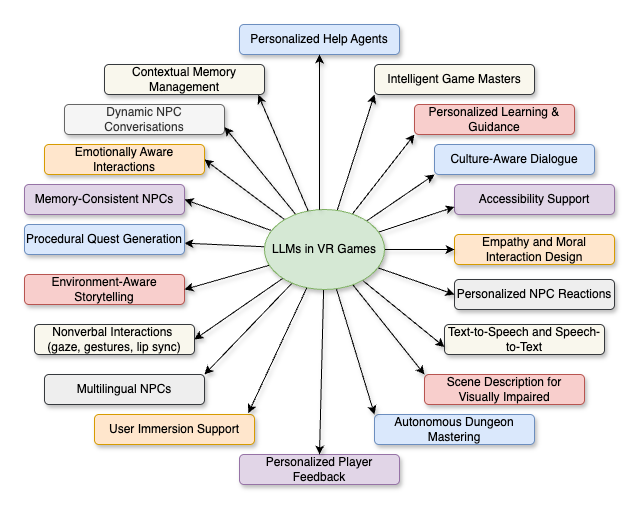}
    \caption{Application areas of Large Language Models in Virtual Reality games}
    \label{fig:llm-usage-vr}
\end{figure}

The applications of LLMs in VR span several key domains, including NPC interaction, game mastering, accessibility, personalization, and ethics. Figure~\ref{fig:llm-usage-vr} provides a visual overview of these areas, many of which are explored in depth throughout this paper to highlight current implementations and future opportunities.

\section{Methodology}

This study employed a structured literature review to examine how Large Language Models (LLMs) are applied in Virtual Reality (VR) games. The goal was to identify current applications, implementation strategies, and key technical and ethical challenges.

\subsection{Literature Search Strategy}

A comprehensive search was carried out across five leading academic databases: \textit{IEEE Xplore}, \textit{ACM Digital Library}, \textbf{Scopus}, \textit{Web of Science}, and \textit{Google Scholar}. These sources were chosen for their extensive coverage of peer-reviewed research in computer science, artificial intelligence (AI), and immersive technologies.

Various search terms were used in different Boolean combinations (AND, OR) to capture a broad and relevant set of studies. To improve precision, we used Boolean logic to combine general terms (e.g., “virtual reality”) with model-specific terms (e.g., “GPT-4”, “conversational agent”), and excluded non-relevant acronym matches. Wherever supported by the database, double quotation marks (e.g., "virtual reality", "NPC dialogue") were used to ensure exact phrase matching and reduce noise in search outputs. 

These search terms targeted the intersection of LLMs and VR across multiple application areas such as NPC interaction, procedural storytelling, accessibility, personalization, and system performance. Table~\ref{tab:search-terms} summarizes the main categories and example search terms used during the database queries.

\begin{table*}[htbp]
  \centering
  \caption{Search Terms and Categories Used in the Literature Review}
  \label{tab:search-terms}
  \setlength{\tabcolsep}{4pt}           
  \renewcommand{\arraystretch}{1.1}     
  \begin{tabular*}{\textwidth}{@{\extracolsep{\fill}}|p{0.23\textwidth}|p{0.72\textwidth}|}
    \hline
    \textbf{Search Term} & \textbf{Example Keywords and Phrases} \\
    \hline\hline
    LLMs in VR Games
      & ``LLMs in VR Games'', ``GPT in VR Games'', ``NPC Dialogues in Virtual Reality'',
        ``LLM‑based conversational agents in VR'' \\
    \hline
    AI Storytelling
      & ``Large Language Models in Gaming'', ``Procedural Storytelling with AI'',
        ``AI‑generated narratives in immersive environments'' \\
    \hline
    NPCs and Game Masters
      & ``AI-driven NPC interactions in VR'', ``Intelligent Game Masters in VR'',
        ``LLMs as Game Masters in VR'', ``AI‑controlled characters in virtual environments'' \\
    \hline
    Player Personalization
      & ``Personalized Player Experience with LLMs'', ``Adaptive narrative systems in VR'',
        ``Behavior-driven dialogue generation'' \\
    \hline
    Performance Issues
      & ``LLM latency in VR games'', ``Real-time response in AI-driven gameplay'',
        ``Computational overhead in immersive environments'' \\
    \hline
    Memory and Ethics
      & ``LLM memory management for virtual agents'', ``Ethical issues in VR AI'',
        ``Long-term interaction in AI systems'' \\
    \hline
    Bias and Safety
      & ``Bias in LLM-based NPCs'', ``Trust and safety in conversational agents'',
        ``Content moderation in AI-driven games'' \\
    \hline
    Immersion
      & ``Immersion and believability in AI-driven VR'', ``User experience with AI NPCs'',
        ``User experiences in VR'' \\
    \hline
    Deployment
      & ``Deployment challenges for LLMs in VR'', ``Scalability of LLMs in VR games'',
        ``Integration of LLMs into game engines'' \\
    \hline
  \end{tabular*}
\end{table*}

Clear inclusion criteria were established to guide the selection of relevant studies. These criteria prioritized recency, relevance to LLM and VR integration, and practical application in gaming or adjacent immersive contexts. Table~\ref{tab:inclusion-criteria} presents the conditions that studies had to meet to be considered in the final review.

\begin{table*}[htbp]
  \centering
  \caption{Inclusion Criteria for Selected Literature}
  \label{tab:inclusion-criteria}
  \setlength{\tabcolsep}{4pt}           
  \renewcommand{\arraystretch}{1.1}     
  \begin{tabular*}{\textwidth}{@{\extracolsep{\fill}}|p{0.15\textwidth}|p{0.82\textwidth}|}
    \hline
    \textbf{Criterion} & \textbf{Description} \\
    \hline\hline
    Time Frame
      & Studies published between 2018 and 2025, covering the period in which transformer‑based LLMs emerged and were first explored in immersive technologies \\
    \hline
    Publication Type
      & Peer‑reviewed journal articles, conference papers, or technical reports with publicly available full texts \\
    \hline
    Topical Relevance
      & Focused on the integration of LLMs into VR environments, with a primary emphasis on gaming contexts \\
    \hline
    Related Applications
      & Papers addressing adjacent VR domains such as education, training, or accessibility using LLMs, provided they offer transferable insights for VR gaming \\
    \hline
  \end{tabular*}
\end{table*}

Alongside the inclusion conditions, exclusion criteria were defined to eliminate irrelevant or low-quality sources. These criteria ensured that only studies with substantial technical contributions and contextual alignment were analyzed. Table~\ref{tab:exclusion-criteria} outlines the reasons for omitting certain works from the review.

\begin{table*}[htbp]
  \centering
  \caption{Exclusion Criteria for Literature Screening}
  \label{tab:exclusion-criteria}
  \setlength{\tabcolsep}{4pt}           
  \renewcommand{\arraystretch}{1.1}     
  \begin{tabular*}{\textwidth}{@{\extracolsep{\fill}}|p{0.15\textwidth}|p{0.82\textwidth}|}
    \hline
    \textbf{Criterion} & \textbf{Description} \\
    \hline\hline
    Language
      & Studies not written in English, to ensure consistency and accessibility during the review process \\
    \hline
    Scope
      & Publications unrelated to either LLMs or VR/immersive environments, or papers where the two technologies were discussed independently without meaningful integration \\
    \hline
    Depth of Contribution
      & Conceptual or commentary papers without technical implementation, empirical evaluation, or practical design frameworks involving LLMs in VR \\
    \hline
  \end{tabular*}
\end{table*}

While the primary focus of the research review was on recent publications, some earlier works were also included, particularly those published before 2015. These were not used mainly to analyze "Key Applications" or "Challenges and Limitations" parts, but served as foundational sources to explain core concepts and historical context in the Background section. 

The initial database queries returned a total of 528 records across five academic databases. After removing duplicates, 422 papers remained. Title-based relevance filtering further reduced the pool to 250 studies, which were then screened based on their abstracts. Of these, 89 papers were selected for full-text review due to their relevance to LLM integration in VR or transferable insights into virtual game design. Ultimately, 62 papers were included in the final literature review.

Due to the limited number of studies focused exclusively on “LLMs in VR games,” the inclusion criteria were purposefully extended to papers addressing related topics, such as LLMs in non-VR games or LLM-powered interactions in VR simulations, which were analyzed to extract applicable insights. A snowballing method was also employed, wherein frequently cited and conceptually central studies were traced backward from reference sections.

\begin{table}[htbp]
\centering
\caption{Summary of Literature Screening and Selection Process}
\label{tab:screening-summary}
\begin{tabular}{|p{0.6\linewidth}|p{0.3\linewidth}|}
\hline
\textbf{Stage} & \textbf{Papers} \\
\hline
Total papers retrieved from databases & 528 \\
\hline
After duplicate removal & 422 \\
\hline
After title-based filtering & 250 \\
\hline
Abstract-screened papers & 250 \\
\hline
Full-text papers reviewed & 89 \\
\hline
Final papers included in review & 62 \\
\hline
\end{tabular}
\end{table}

\begin{figure}[htbp]
    \centering
    \includegraphics[width=0.5\textwidth]{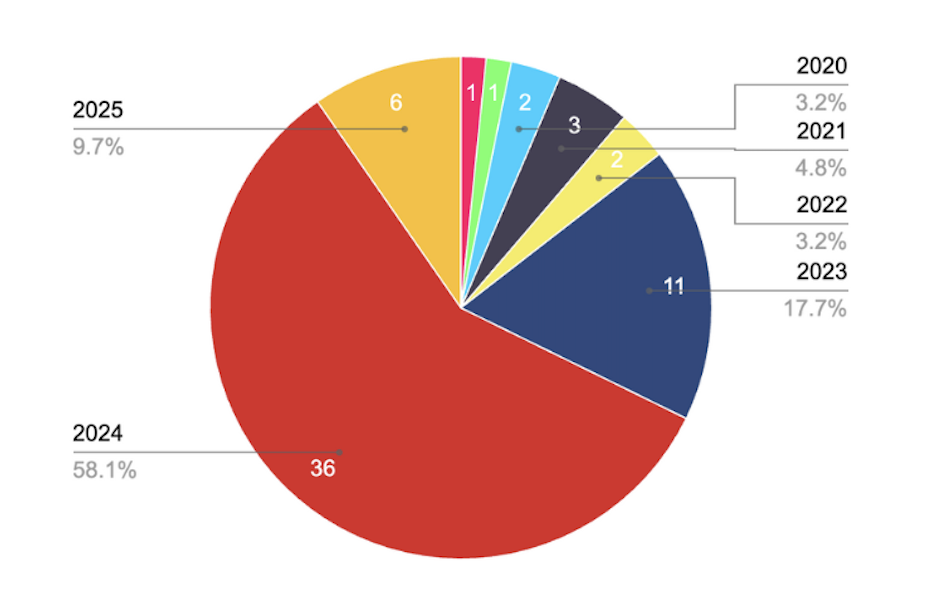}
    \caption{Distribution of reviewed papers by publication year.}
    \label{fig:year-distribution}
\end{figure}

The distribution of reviewed papers by publication year reflects a sharp increase in research interest, especially after 2023, peaking in 2024 and 2025 (see Figure~\ref{fig:year-distribution}).

Following the full-text analysis of 62 selected studies, a thematic categorization was conducted to structure the findings and identify primary research directions. The classification framework was based on the scope of each study and the recurring concepts in the literature. Studies were grouped according to their primary application domains, key findings, and the challenges they addressed. Two main thematic categories emerged:

\begin{itemize}
    \item {Key Applications of LLMs in VR Games}
    \item {Challenges and Limitations in Implementation}
\end{itemize}

We divided these categories into subcategories based on common research objectives, technical approaches, and experimental setups. The classification is aligned with the structure of the literature review section in this paper, which includes:

\begin{enumerate}
    \item \textbf{Dynamic NPC Interactions and Emotional Intelligence:} Studies that implement LLMs to enhance the realism and responsiveness of NPC dialogue and behavior.
    \item \textbf{Procedural Storytelling and Narrative Generation:} Works on generating adaptive, personalized narratives using LLMs, often in role-playing games or branching story environments.
    \item \textbf{Intelligent Game Masters and Adaptive Systems:} Research using LLMs for autonomous scene control, improvisational gameplay, and dynamic environment management.
    \item \textbf{Personalized Player Experiences:} Papers discussing systems that tailor content, difficulty, or narrative tone based on player preferences and interaction history.
    \item \textbf{Accessibility and Inclusivity:} Studies that leverage LLMs for real-time translation, multimodal interaction, and interface personalization to support diverse player groups.
    \item \textbf{Computational and Performance Constraints:} Research that addresses latency, memory management, and computational costs in real-time LLM deployment within VR games.
    \item \textbf{Ethical and Safety Considerations:} Papers exploring content moderation, bias mitigation, and privacy in AI-driven VR applications.
    \item \textbf{User Experience and Immersion Issues:} Studies examining the perceptual quality of interactions with LLM-powered NPCs, focusing on believability, consistency, and emotional engagement.
    \item \textbf{Scalability and Deployment Barriers:} Contributions discussing challenges in bringing experimental LLM+VR systems into real-world, multi-user, or commercial settings.
\end{enumerate}

This taxonomy provides a structured synthesis of current research and helps identify emerging opportunities and gaps in integrating LLMs with VR games. It also enables comparative evaluation of different technical and design strategies.

\begin{figure}[htbp]
    \centering
    \includegraphics[width=1\linewidth]{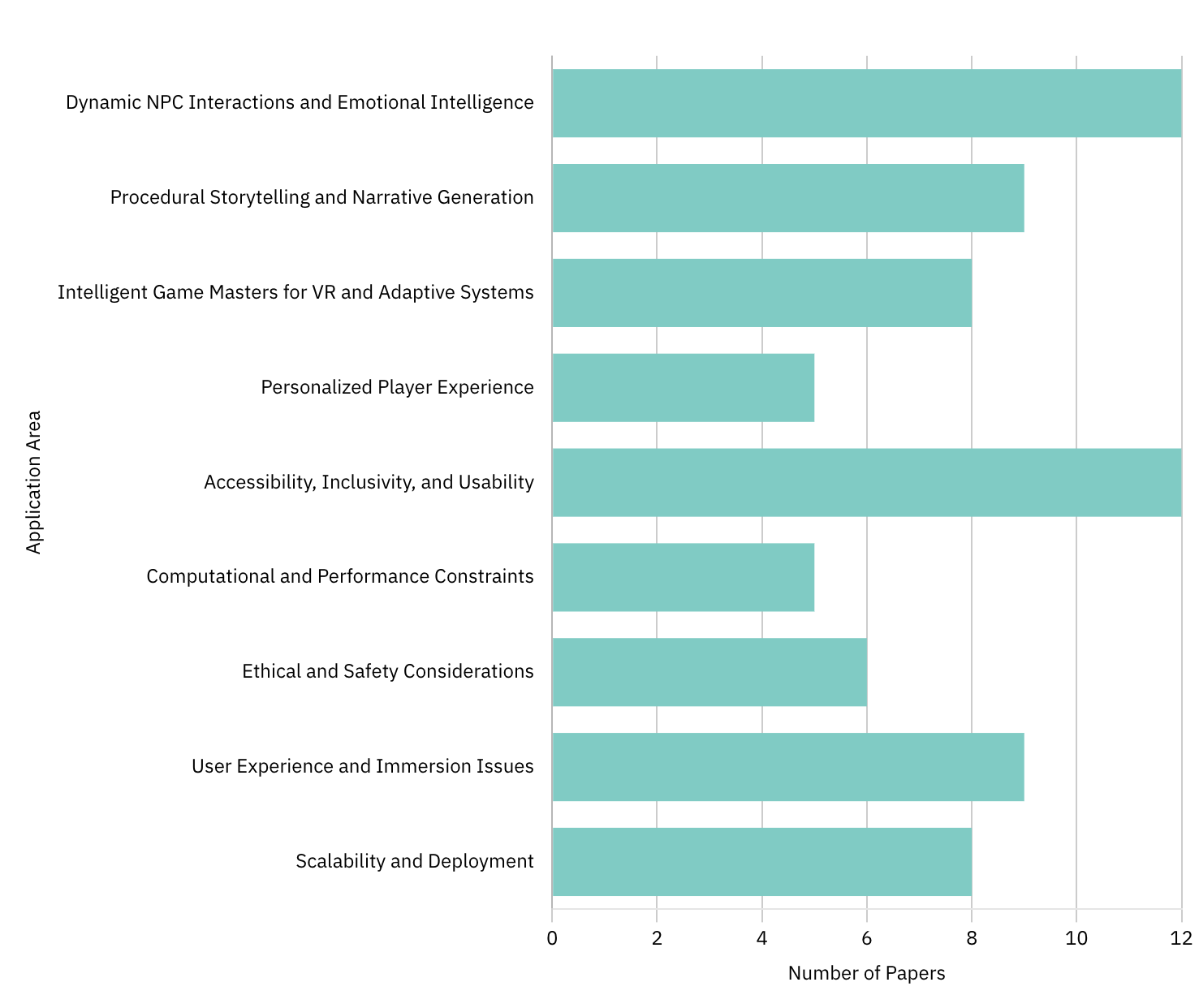}
    \caption{Categorization of reviewed papers based on key application areas. Note that one paper may be relevant to multiple categories.}
    \label{fig:category-distribution}
\end{figure}

Figure~\ref{fig:category-distribution} illustrates the approximate number of papers associated with each key application area to visualize the distribution of reviewed studies across these categories. Note that individual studies may span multiple categories if they contribute meaningfully to more than one domain.

\section{Key Applications of LLMs in VR Games}

As the convergence of Large Language Models (LLMs) and Virtual Reality (VR) advances, various impactful use cases have emerged across the VR gaming landscape.

\subsection{Dynamic NPC Interactions and Emotional Intelligence}

Recent advancements in LLMs have enabled the development of more responsive and emotionally intelligent NPCs within VR environments. Modern NPCs move away from fixed, pre-programmed behaviors because they now use LLMs to generate dynamic conversational abilities, emotional expression, and adaptive interaction strategies \cite{sweetser2024llmsgames}. This section combines essential research findings to show how LLMs improve character behavior and dialogue delivery in virtual reality environments.

\paragraph*{Emotionally Expressive NPCs with LLMs}
The use of LLMs such as GPT-3.5 and GPT-4 to give NPCs facial expressions, gestures, and emotionally human-like dialogue is on the rise.

Normoyle et al., \cite{normoyle2024using} used GPT-3.5 to create facial expressions, body movements, and lip-syncing for game characters based on what was being said. They used the Facial Action Coding System (FACS) and Laban Movement Analysis (LMA) to guide the animations. Their study was done in a 3D point-and-click game, not real-time VR. One limitation was that small changes in prompts could cause inconsistent animations, a known issue with LLMs. Despite this, the study shows that LLMs can automatically help generate emotional character behavior, saving time and making interactions feel more lifelike.

Building on this, Marincioni et al. \cite{marincioni2024effect} studied how LLMs could assign emotions like Happy, Sad, Angry, or Neutral to NPCs in a mystery game, and how these emotions affected players. Interestingly, players often reacted positively even to negative emotions, such as gratitude toward angry NPCs. This reveals how emotionally expressive NPCs can create complex psychological responses. The study shows that giving NPCs emotional depth using LLMs can greatly enhance immersion and shape the overall gameplay experience.

\paragraph*{Personality and Conversational Naturalism}

Consistency in NPC personality is essential for believable and immersive interactions. Hasani and Udjaja \cite{Hasani2021DynamicDialogue} proposed an early framework combining generative dialogue, emotional cues, and multimodal interaction to support personality-consistent, context-aware responses. Building on this, Zhu et al., \cite{zhu2023freeform} found that users retained more information and felt more immersed when engaging with human-like avatars than abstract ones. Similarly, Tonini’s international study \cite{tonini2024talk} showed that voice-driven AI NPCs enhanced user experience through emotionally engaging and polite communication, though issues like latency and limited memory reduced sustained immersion.

\paragraph*{Memory, Consistency, and Long-Term Interaction}

Ensuring consistent dialogue over time remains a key challenge in AI-driven NPC design. Zheng et al., \cite{zheng2024memoryrepository} proposed a dual-memory system (MemoryRepository) that mimics human-like forgetting and summarization, allowing NPCs to recall both recent and long-term interactions. Tested with models like GPT-4, GPT-3.5, and ChatGLM, the system improved dialogue continuity, engagement, and immersion. In a related approach, Jahangiri et al., \cite{jahangiri2024balancing} focused on optimizing performance by combining LLMs with Pursuit Learning Automata (PLA). Their hybrid system enabled faster responses and dynamically adjusted dialogue tone to match player preferences, balancing emotional richness with real-time scalability.

\paragraph*{Multimodal and Nonverbal Interactions}

LLMs also support multimodal NPC interactions by combining voice, gaze, and gesture, making characters more lifelike and responsive. Players tend to prefer NPCs that recognize physical gestures, such as waving or nodding, and provide real-time feedback through cues like lip-syncing and state lights. Yin and Xiao \cite{yin2024press, krupka2017toward} analyzed 47 VR games and found that physical actions significantly enhance immersion. Players expected NPCs to respond to proximity, gestures, and eye contact, making them feel more aware and reactive. Maslych et al. \cite{maslych2025takeaways} further emphasized the role of feedback cues like state lights, gaze, and facial expressions during conversations, which increased users’ trust and engagement. Even simple indicators during system response times, such as loading bars, reassured users that the NPC was actively processing their input. Sissler \cite{sissler2024enhancing} developed an open-source Unity framework using GPT-3.5 that integrates voice, gestures, and animated facial expressions. The study showed that synchronized multimodal responses improved NPC believability and helped players feel heard. These findings highlight that real-time multimodal feedback is key to creating immersive and socially engaging LLM-driven NPCs in VR.

Across the reviewed studies, LLMs such as GPT-3.5 and GPT-4 are widely adopted to create emotionally expressive, socially aware NPCs capable of real-time dialogue. While many systems simulate facial expressions, gestures, and vocal affect, most were tested outside of fully immersive VR settings. Research consistently highlights the importance of consistent personality, emotional depth, and long-term memory in maintaining user immersion. However, limitations remain in sustaining coherence over extended interactions and minimizing latency in real-time environments. The integration of multimodal cues—voice, gaze, gesture—has proven especially effective in enhancing believability and player engagement.

\subsection{Procedural Storytelling and Narrative Generation}

LLMs are changing how games tell stories by automatically creating quests, dialogues, scenes, and storylines that adjust to player actions and preferences. This flexibility shapes players' experiences, making gameplay more immersive and personalized. Recent research has explored ways LLMs support storytelling, such as generating new story paths, building dynamic quests, and adapting to the game’s context. These studies show that LLMs can improve interactive storytelling, though challenges with consistency and coherence remain.

One study used GPT-4-powered NPCs in an interactive fiction game where players could speak freely instead of choosing from pre-written options. This led to unexpected storylines and character relationships that the designers had not planned—players who liked exploring enjoyed this freedom to create complex and personal narratives. However, the system sometimes repeated itself or gave inconsistent replies due to memory limitations \cite{peng2024player}. Another project, PANGeA, combined LLMs with branching logic to create quests and dialogue in a turn-based RPG. The game world changed based on player decisions, leading to unique and replayable stories. While this approach gave players more variety, it sometimes produced plot inconsistencies, especially during long play sessions \cite{buongiorno2024pangea}.

\paragraph*{Procedural Quest and Dialogue Generation}

LLMs have been widely explored for generating quests and dialogues in role-playing games (RPGs). One study fine-tuned GPT-2 using 978 quests from existing games, resulting in a model called Quest-GPT-2. This system produced more varied and creative quests than traditional retrieval-based methods, and human evaluators found that about 20\% of the generated quests were usable without significant edits. However, the model struggled with coherence, often creating quests with unclear goals or inconsistent character relationships, especially in multi-step storylines \cite{vartinen2024generating}.

Another approach used knowledge graphs alongside LLMs to improve coherence and relevance. The system combined information about characters, history, and player choices to generate quests and dialogues aligned better with the game world. Human reviewers rated these quests higher for fluency and logical consistency than standard LLM output. Still, the method faced challenges with memory retention and maintaining story consistency over longer play sessions \cite{ashby2023personalized}.

Further development came through a persona-based framework that used LLMs to generate consistent character dialogues across different scenes. This system used “persona cards” to define character traits and “scene cards” to give context, which helped LLMs maintain each NPC’s personality over time. Combining prompt engineering and fine-tuning, even smaller LLMs (with around 7 billion parameters) could produce high-quality, personality-rich dialogues \cite{bingli2024towards, yang2023baichuan2}.

\paragraph*{Scene, Context, and Environment-Aware Storytelling}

Recent studies have focused on how LLMs can generate narratives and dialogue that adapt to in-game environments and context. One notable system, SceneCraft, used GPT-4 to create interactive scenes and cutscenes by combining predefined templates with probabilistic variation. Developers could define scene structures, and the system would expand them into coherent story events. While the generated scenes were engaging and consistent in individual instances, maintaining character and world consistency across multiple scenes remained a challenge \cite{kumaran2023scenecraft}.

Radež and Bohak \cite{radez2024integrating} introduced a system that enabled NPCs to generate dialogue based on their awareness of the game environment. Using panoramic image capture and semantic segmentation, NPCs could reference nearby objects and spatial relationships, creating more believable and immersive interactions. Players appreciated the added realism, but the system’s high computational demands limited its application in real-time VR settings.

Li et al., \cite{li2025environmentaware} proposed a schema-based prompting method for GPT-4 Turbo agents to handle spatial interactions in VR, such as pointing, grabbing, or navigating scenes. Their system generated dialogue based on environmental cues, object properties, and user actions. The agents were tested in various role-play scenarios and showed practical spatial reasoning and responsiveness. However, they sometimes hallucinated object references and struggled in more complex environments.

Earlier work by Hämäläinen et al., \cite{hamalainen2019creative} demonstrated a system that adapted NPC dialogue in Fallout 4 based on gameplay variables like health or quest progress. Instead of generating new lines, the system rephrased existing ones to better fit the player's current state. While effective for personalization, it lacked memory of past interactions and could not support dynamic long-term conversations.

The reviewed studies demonstrate that LLMs can significantly enhance procedural storytelling by generating dynamic quests, dialogues, and scene-aware narratives. Techniques such as fine-tuning, prompt engineering, and knowledge graph integration help maintain coherence and character consistency, though challenges persist with memory retention and logical continuity across extended sessions. Systems like SceneCraft and schema-based prompting show promise in generating context-aware scenes and spatially grounded dialogue, yet often face scalability limitations in real-time VR settings. Overall, while LLMs expand narrative flexibility and personalization, consistent world-building and long-term dialogue coherence remain open challenges for future development.

\subsection{Intelligent Game Masters for VR and Adaptive Systems}

Recent advancements in LLMs have made it possible to create AI Dungeon Masters (DMs) capable of managing player-driven narratives and improvising gameplay in real time. Several studies show that AI DMs can take over key storytelling responsibilities typically handled by human game masters, enhancing the role-playing experience. For instance, ChatGPT has been explored as a DM for tabletop role-playing games (TTRPGs) like Dungeons \& Dragons. It was able to generate coherent narratives and respond to player input dynamically. However, the study also noted limitations, such as delayed responses and limited emotional engagement, which affected player immersion \cite{triyason2023exploring}.

To support novice game masters, Kelly et al. expanded a tool called Shoelace by adding LLM-based dialogue suggestions and information retrieval. This helped users manage scenes and improvise more effectively, especially beginners \cite{kelly2023towards}.

Another study focused on improving AI DMs by integrating function calling into LLMs. In the context of Jim Henson’s Labyrinth: The Adventure Game, the system used two types of functions: one for simulating dice rolls and another for updating the game state. Combining both functions led to more consistent and engaging storytelling, as the AI could better follow game rules and handle random events \cite{song2024labyrinth}.

Research has also looked into how the personality of AI game masters affects players. Findings show that players respond more positively to friendly and cooperative AI DMs, suggesting that the tone and demeanor of the AI can influence both gameplay and player emotions \cite{you2024dungeons}.

\paragraph*{Adaptive and Interactive AI Systems in VR Games}

LLMs are now used in tabletop games and as adaptive assistants and world managers in dynamic virtual environments. One early example involved a GPT-based voice assistant in a low-cost VR escape room, which provided hints and story cues based on the game’s context. While this improved gameplay through adaptive responses, it faced challenges such as response delays and limited real-time flexibility \cite{rychert2024integrating}.

Beyond desktop and cloud-based systems, some studies explored lightweight mobile VR applications. For example, Khan et al. created a multiplayer VR carrom game where players competed against an AI opponent using Bluetooth controls and first-person vision. This demonstrated an early attempt to integrate AI-driven decision-making into mobile VR platforms \cite{khan2018carromvr}.

In more complex scenarios, LLMs have been used to control multi-agent teams in adversarial search-and-rescue games. These AI agents outperformed traditional strategic planning and opponent modeling models by using advanced prompting methods such as Zero-shot Chain-of-Thought (CoT) and iterative cue-based learning \cite{li2024seek}.

The LLMR framework furthers this by offering a modular system for managing interactive virtual worlds. It uses multiple GPT-based modules to handle scene understanding, task planning, and debugging. This setup enables real-time 3D scene creation with fewer errors and greater coherence than using a single LLM alone \cite{delaTorre2024llmr}. Together, these studies show how LLMs can support adaptive, responsive, and intelligent control of VR game environments.

LLMs are increasingly leveraged to function as intelligent game masters and adaptive agents in VR, capable of facilitating improvisational storytelling, rule-based decision-making, and multi-agent coordination. Studies show that ChatGPT and similar models can effectively manage narrative flow and simulate dynamic events, though response delays and limited emotional depth still affect immersion. Integrations with function calling, tone customization, and scene management tools like Shoelace and LLMR have improved coherence and flexibility. However, most applications remain experimental or limited to lightweight systems, highlighting the need for further optimization for real-time, large-scale VR environments.

\subsection{Personalized Player Experience}

LLMs offer new ways to personalize gameplay in VR by enabling adaptive dialogue, emotional feedback, and context-aware storytelling. Personalization now goes beyond adjusting difficulty or settings—it involves creating AI-driven agents that can understand, remember, and respond to players in socially intelligent and emotionally engaging ways. While some aspects have been discussed earlier, this section focuses on broader strategies such as creativity support, player modeling, and emotionally tailored narratives.

One primary use of LLMs in personalization is enabling natural conversations between players and virtual agents. Studies have shown that players remember more and engage longer when interacting with LLM-powered avatars, especially when the agent remembers previous interactions and maintains a human-like personality \cite{zhu2023freeform, tonini2024talk}.

A promising direction involves using LLMs to support player creativity. Lin et al., \cite{lin2024application} developed a VR brainstorming system where ChatGPT-powered NPCs acted as creative partners. These assistants offered voice suggestions, summarized discussions, and retrieved relevant information in real time. The system encouraged divergent thinking and collaborative idea generation by understanding the ongoing conversation, turning the AI into a co-creator rather than just a tool.

Tucek \cite{tucek2024enhancing} explored emotionally personalized storytelling, where NPCs adapted to each player's social identity, emotional state, and choices. These emotionally aware agents used LLMs to generate real-time dialogue aligned with the player’s perspective, aiming to foster empathy and deeper narrative engagement.

Other studies focused on personalizing VR experiences through familiarity. Guo et al., \cite{guo2023whos} found that players responded better to NPCs that looked or sounded familiar. In exergames, avatars that reflected users’ preferences improved enjoyment and performance, particularly for more self-conscious players. These results show that even simple visual or auditory customization can enhance user experience in measurable ways.

LLMs are enabling highly personalized VR experiences by supporting emotionally aware, conversationally adaptive, and context-sensitive virtual agents. Studies highlight that memory retention, personality continuity, and co-creative dialogue enhance user engagement and satisfaction. Creative assistance, emotionally aligned storytelling, and familiarity-based customization have been shown to foster empathy, enjoyment, and performance. However, most implementations remain small-scale or experimental, and sustaining long-term personalization in complex, dynamic environments remains a challenge for future work.

\subsection{Accessibility, Inclusivity, and Usability}

LLMs are increasingly used in VR environments to improve accessibility, inclusivity, and usability. Their integration supports the development of adaptive systems that address diverse user needs, including individuals with disabilities, language differences, or limited technological experience. This section explores how LLMs enhance VR through multimodal assistance, personalized dialogue, and culturally sensitive design.

\paragraph*{Multimodal and Sensory Accessibility}
One key advantage of LLMs is their ability to generate natural language explanations for users with sensory limitations. Multimodal models like GPT-4V enable scene descriptions via text-to-speech, helping visually impaired users navigate virtual spaces. For example, EnVisionVR interprets 360-degree scenes to provide real-time audio feedback on spatial layouts and object locations \cite{chen2025envisionvr}.

LLMs also support inclusive design by enabling dynamic, user-adaptive interactions. Bozkir et al., \cite{bozkir2024embedding} argue that LLM-powered NPCs can adjust to different user needs through prompt engineering and fine-tuning, offering more personalized and equitable experiences than static, pre-scripted agents.

\paragraph*{VR Games for Mental Health and Well-being}
Baghaei et al. \cite{baghaei2020ivr} conducted a design-driven study exploring how individualized virtual reality (iVR) environments could enhance mental health outcomes, particularly among young people aged 18–25. Drawing on prior work by Falconer et al. \cite{falconer2016selfcompassion}, they implemented a VR experience aimed at increasing self-compassion as a pathway to alleviating depressive symptoms. Participants could personalize key aspects of the virtual environment, including the avatar, therapeutic setting, and avatar behaviors. The study found that such personalized experiences were perceived as more meaningful, emotionally engaging, and safer than standardized VR therapy. Personalization—especially when tailored to the user’s identity, emotional state, and goals—was shown to enhance users’ motivation and sense of connection. These findings support the potential of iVR to provide scalable, user-centered mental health interventions.

Baghaei et al. \cite{baghaei2021vrmentalhealth} conducted a scoping review of 34 studies that used VR to treat depression and anxiety. Their findings indicate that the majority of included studies reported positive therapeutic outcomes when VR was used as part of a treatment strategy. Notably, nine of these studies applied cognitive behavioral therapy (CBT) within or alongside VR environments, all of which reported a reduction in symptoms. The review highlighted that VR-based CBT was not only effective but also practical for clinicians, allowing for standardized delivery, repeatability, and increased patient engagement. The authors concluded that VR shows strong potential for structured mental health interventions, especially when it leverages immersive interaction and controlled exposure through techniques like VRET (Virtual Exposure Therapy).

Chitale et al. \cite{chitale2022assessment} presented a scoping review focused on the use of both video games and VR for assessing anxiety and depression. Out of 4566 records initially screened, 10 studies were included, split evenly between VR and videogame-based approaches. An important trend noted in the findings was that studies on anxiety predominantly used VR, while those on depression leaned toward traditional video games. A few studies incorporated machine learning techniques, and only two were clinical trials. Most studies yielded encouraging outcomes, suggesting that both modalities could be useful tools for assessment. However, the authors stressed the limited availability of high-quality clinical evidence and recommended closer collaboration with mental health professionals to ensure safety and privacy in future development.

Given their adaptability, conversational fluency, and capacity for emotionally responsive interaction, LLMs hold strong potential to enhance these therapeutic VR experiences, particularly by supporting personalized narratives, mood-aware guidance, and dynamic user engagement in mental health contexts \cite{berrezueta2025therapeutic}.

\paragraph*{Social and Educational Inclusion}
Beyond accessibility, LLMs have shown potential in fostering inclusion for individuals with diverse cognitive and learning needs. Li et al. \cite{li2024exploring} implemented LLM-based chatbots within VR job interview simulations designed for autistic users. These virtual agents offered personalized, voice-based feedback in low-pressure, repeatable environments. The structured yet flexible format helped users build communication skills while maintaining a sense of psychological safety, an essential aspect for neurodivergent learners navigating real-world scenarios.

Similarly, Voultsiou et al. \cite{voultsiou2025aiinclusive} explored the use of LLM-powered assistants in VR learning environments tailored to students with special educational needs, including autism. Their findings indicate that AI-driven guidance enhanced learner engagement and comprehension, especially when combined with multimodal inputs like visual cues or simplified language. However, they also observed that the current systems often lack sufficient depth in personalization and struggle to maintain long-term contextual awareness, which limits their effectiveness across extended educational sessions.

Additional studies on usability show that integrating natural input modalities such as hand tracking further improves interaction quality. Geetha et al. and Krupka et al. \cite{geetha2024human, krupka2017toward} emphasize that users—particularly those unfamiliar with game controllers—benefit from gesture-based systems that provide intuitive, real-time feedback. These affordances make VR more approachable for a broader range of users, from children with learning difficulties to older adults or those with motor impairments.

\paragraph*{Cultural and Contextual Usability}
Cultural usability in VR is gaining attention as a means to make virtual environments more relatable and engaging for diverse user groups. LLMs, with their capacity for dynamic language generation and contextual adaptation, are increasingly being used to enhance cultural relevance in VR narratives. Lau et al., \cite{lau2024scottishvr} explored this in a Scottish curling game, where NPCs used culturally appropriate language and expressions. Participants reported that the familiar tone and regional references significantly improved their sense of presence and emotional connection to the experience, demonstrating that localized dialogue—powered by LLMs—can heighten user engagement in culturally specific scenarios.

Similarly, Subandi et al. \cite{subandi2022sasirangan} developed a VR shopping simulation designed to preserve and promote Indonesian textile heritage. Users were guided by LLM-enabled NPCs through the traditional Sasirangan fabric-making process. The agents not only narrated historical context but also responded to questions, allowing for interactive exploration. This use of LLMs for cultural storytelling helped users engage with intangible cultural knowledge in a personalized and immersive manner, suggesting new possibilities for cultural preservation and education through interactive AI.

In broader educational and heritage applications, LLMs have been used to power intelligent virtual tutors capable of delivering contextualized instruction. For instance, Ayre et al., \cite{ayre2023laboratory} created a GPT-4-based assistant for a virtual chemistry lab. This tutor provided step-by-step instructions and real-time support tailored to users' actions, effectively acting as a dynamic guide. Users reported increased understanding and autonomy, attributing it to the tutor’s ability to interpret the learning context and offer personalized feedback.

Together, these studies show how LLMs enhance cultural and contextual usability in VR by offering adaptive, linguistically nuanced, and locally grounded interactions. This not only improves accessibility for diverse populations but also enriches the educational and emotional value of VR content.

LLMs improve accessibility, inclusivity, and usability in VR through adaptive multimodal and context-aware interactions. Users with sensory limitations benefit from LLMs because they provide real-time audio guidance, adaptive dialogue, and intuitive interfaces that enhance VR navigation and responsiveness. The application of LLMs shows great promise for therapeutic interventions because they create emotionally responsive and personalized VR scenarios that benefit patients undergoing anxiety, depression, and PTSD treatments.

The implementation of LLM-based assistants in educational and social training environments has enhanced communication abilities and learning outcomes and user confidence for autism and cognitive difference users, while gesture-based inputs make the system more accessible to new users. The implementation of LLMs with cultural and contextual adaptations through localized narratives and intelligent tutoring systems demonstrates their ability to enhance user engagement in heritage, educational, and commercial VR experiences. Future research needs to resolve essential challenges, which include deep personalization capabilities, sustained memory retention, and real-time system performance limitations.

\section{Challenges and Limitations}

While LLMs offer transformative possibilities for VR games, their integration introduces significant technical, ethical, and usability challenges. 

\subsection{Computational and Performance Constraints}

Although many studies explore how language models can enhance interactive systems, most have not yet been tested in real immersive VR settings. Instead, evaluations are often done on desktop platforms, where performance issues like latency, motion tracking, and multimodal input are less demanding. This creates a gap in our understanding of how LLMs behave under real-time, resource-intensive VR conditions, where delays or instability can negatively impact user experience.

Several studies report that using LLMs in VR requires substantial computational resources. For example, Maslych et al. \cite{maslych2025takeaways} found that even with local deployment and optimizations like automatic speech recognition (ASR), text-to-speech (TTS), and behavior-state modeling, response times averaged 3.2 seconds—too slow for real-time interaction. Running LLMs locally instead of via cloud APIs helps reduce delay, while behavior-state modeling, which defines agent states like listening or speaking, supports more synchronized interactions. Still, these methods don’t fully solve latency issues in VR.

Jahangiri and Rahmani \cite{jahangiri2024balancing} observed longer delays—over 20 seconds—in LLM-based NPC systems. They combined LLMs with Pursuit Learning Automata (PLA) to address this, creating a hybrid setup that reduced response time to under one second. While promising, this approach still requires careful tuning and is difficult to generalize across different VR environments.

Memory limitations are another critical barrier. As Zheng et al. \cite{zheng2024memoryrepository} point out, LLMs struggle to maintain consistent conversations over time. Their MemoryRepository system mimics human memory by summarizing past interactions, helping sustain dialogue coherence. However, this adds processing demands, which may not scale well in complex or multi-character VR scenarios.

Making NPCs aware of their environment adds more complexity. Radež and Bohak \cite{radez2024integrating} used image capture and semantic segmentation to let NPCs reference objects and spaces around them. While this improves realism, the real-time processing it requires is complex to achieve on typical consumer VR hardware, making widespread use difficult without sacrificing performance.

Sissler \cite{sissler2024enhancing} demonstrated improved NPC dialogue using GPT-3.5 in Unity, but delays from REST API calls still reduced immersion. The study recommends switching to stream-based architectures for faster response. It also highlights the need for expert prompt engineering to achieve natural conversations. Significantly, while LLMs enhance language-based interactions, key NPC behaviors—like movement and planning—still depend on traditional scripting, limiting full autonomy.

LLMs offer rich linguistic and expressive capabilities for VR, but their integration into real-time immersive environments faces significant technical challenges, including latency, memory constraints, and computational overhead. Even with local deployment and optimizations, current systems often fall short of the responsiveness needed for seamless interaction, with some reporting delays up to 20 seconds. Approaches such as hybrid architectures, memory repositories, behavior-state modeling, and stream-based communication offer partial improvements, yet scalability remains limited. To bridge the gap between expressive AI and immersive VR design, future work should focus on lightweight models, edge computing, and tighter integration with traditional game logic.

\subsection{Ethical and Safety Considerations}

Integrating LLMs into VR games has led to rapid progress in interactive storytelling, emotional expression, and intelligent gameplay. However, as these systems become more adaptive and human-like, a critical question arises: Can we trust AI agents that learn and respond to us in real time? While the potential for immersive and personalized experiences is exciting, it also brings serious ethical and safety concerns, particularly in VR environments where users may build emotional bonds with AI characters \cite{sobchyshak2025pushing}.

One primary concern is privacy. VR systems collect detailed biometric and behavioral data, such as gaze, voice, movement patterns, and emotional cues. Unlike traditional web apps, this data is continuous and fine-grained. Garrido et al., \cite{garrido2023sok} showed that just a few minutes of telemetry data, like eye tracking and EEG signals, can reveal private information such as a user’s gender, income level, or emotional state. These findings emphasize the need for stricter safeguards around data use in immersive settings. One design solution might be to implement consent-aware logging mechanisms, which inform users of what data is being collected and allow them to enable or disable specific data tracking modalities.

In addition to privacy, LLM-generated content raises risks of bias and misinformation. Yang et al., \cite{yang2024gpt} found that GPT-based agents in mixed-initiative gameplay (MIG) can produce biased or misleading stories, especially problematic in educational or therapeutic games. When LLMs are used without proper moderation, they can unintentionally reinforce harmful stereotypes or distort learning outcomes. Proactive bias mitigation can be addressed at the prompt level through controlled prompt engineering and content filtering techniques tailored to sensitive domains such as education or therapy.

Waghale et al., \cite{waghale2024ai} also warn that LLMs can introduce unfairness into gameplay, especially in multiplayer or competitive environments. Bias in training data or algorithm design may lead to advantages or disadvantages for specific player groups. Procedural content generation using big data can unintentionally reinforce cultural or gender stereotypes. At the same time, using sensitive data to personalize gameplay raises significant privacy concerns.

Another challenge is the emotional impact of AI-driven empathy systems. Tucek \cite{tucek2024enhancing} showed that emotionally responsive digital characters behave unpredictably or generate inappropriate content; they can harm user trust or reinforce negative perceptions, especially when the goal is to foster empathy toward marginalized communities. To reduce user confusion or mistrust, transparent AI feedback systems can be used—for instance, by showing visual indicators when an NPC is adapting its behavior in real time.

Tanksale \cite{tanksale2023web3d} adds that LLMs used in immersive Web3D environments pose additional risks when combining real-time personalization with procedural generation. Without oversight, these systems may create biased or culturally insensitive content, especially when trained on unfiltered internet data.

Finally, as Damianova \cite{damianova2025seriousvr} emphasizes, ethical considerations must be built into the design process, not added after deployment. Developers should take responsibility for ethical practices by integrating fairness, inclusivity, and safety principles throughout the design cycle.

These studies highlight the urgent need for responsible AI design, content moderation, and strong privacy protections in VR. Without clear ethical safeguards, the line between helpful personalization and harmful manipulation becomes dangerously thin.

As LLMs bring emotional intelligence and real-time responsiveness into VR games, they also introduce significant ethical and safety risks. Studies consistently show that privacy concerns, algorithmic bias, and unpredictable emotional impacts are not hypothetical—they are already emerging in practice. While adaptive AI systems enhance personalization, they also risk reinforcing harmful stereotypes or manipulating user behavior without clear consent. Addressing these concerns requires embedding fairness, transparency, and safety protocols into every stage of design and deployment, particularly as immersive AI interactions grow more lifelike and emotionally persuasive.

\subsection{User Experience and Immersion Issues}

Creating virtual characters that feel truly lifelike remains one of the biggest challenges in VR game design, especially when using LLMs for NPC dialogue. At the same time, these models can generate fluent and responsive language, which alone does not guarantee engaging or believable interactions in immersive environments. Research shows that the biggest obstacles to user experience are unnatural reactions, inconsistent personality, limited conversational structure, and memory lapses over time.

Maslych et al., \cite{maslych2025takeaways} conducted a pilot study revealing low realism scores (3.12 out of 7) for LLM-driven avatars in task-based VR scenarios. The main issues were minimal animations, limited to basic lip-sync and head movement, which broke immersion. Participants noted that adding facial expressions, idle behaviors, and body motion could improve believability. Visual feedback cues, such as state lights and loading indicators, also played a key role in maintaining user trust by signaling that the avatar was actively listening or processing input.

Tonini’s international study on voice-based VR gameplay highlighted similar issues. While players appreciated LLM-powered NPCs' emotional tone and responsiveness, they also found conversations repetitive and sometimes generic. The lack of dialogue variety and slow reaction times made interactions feel scripted rather than natural. Players enjoyed the freedom of open voice interaction, but the underlying AI often failed to sustain flexible, emotionally rich conversations over time \cite{tonini2024talk}.

A mixed-reality study showed that human-like avatars significantly improved memory retention and immersion compared to symbolic or abstract characters \cite{zhu2023freeform}. This suggests that avatar design, specifically realism, expressiveness, and embodiment, is critical for building emotional user connections. However, delivering this level of engagement requires more than fluent speech. Multimodal feedback, personality modeling, and memory-aware systems must work together to create believable and responsive virtual characters \cite{brito2025integrating}.

Narrative consistency is another primary concern, especially in emergent gameplay. Peng et al., \cite{peng2024player} showed that while players can freely co-create stories with LLM-driven characters, this often leads to fragmented or inconsistent plotlines. When systems fail to remember player actions or goals, the story can feel disjointed and lose its emotional impact.

These findings show that while LLM-based NPCs can deliver moments of intense immersion, they still struggle with sustaining realistic, emotionally coherent, and context-aware interactions. To address this, future work must improve animation, clarity of feedback, narrative memory, and long-term emotional engagement.

Another significant barrier to immersive experience in VR is cybersickness—a form of motion-induced discomfort that affects many users, particularly during fast-paced or unstructured gameplay. It is usually like a physiological response marked as nausea, disorientation, or dizziness. To address this issue, the literature offers a variety of techniques designed to detect and reduce the severity and frequency of cybersickness symptoms \cite{ang2023reduction}. 

Physiological signal analysis has shown great promise for cybersickness detection. Islam et al. \cite{islam2020cybersickness} proposed a deep learning-based method that uses heart rate, breathing rate, heart rate variability, and galvanic skin response to automatically detect and predict cybersickness severity. Their simplified CNN-LSTM model achieved 97.44\% accuracy for current state detection and 87.38\% for predicting future symptoms, outperforming traditional classifiers. This method provides a robust, real-time solution by leveraging subtle physiological changes that correlate with user-reported discomfort. 

In addition to physiological signal analysis approaches, Monteiro et al. \cite{monteiro2021trajectory} demonstrated that trajectory compression rate can also be used as a marker to identify cybersickness during VR gameplay. The authors found a clear correlation between variations in compression rate and users’ Discomfort Scores, indicating that changes in movement patterns—such as increased rotation or erratic navigation—are linked to higher levels of sickness. A simple neural network model using compression rate and its variation as input was able to accurately predict whether discomfort would increase or decrease over time.

Furthermore, Wang et al. \cite{wang2023realtime} presented a novel method for predicting simulator sickness (SS) in real time using only in-game character movement and eye-tracking data, without the need for expensive or external physiological sensors. The authors trained a long short-term memory (LSTM) neural network on data collected from three VR games and achieved an SS prediction accuracy of 83.4\% for players with high sensitivity to SS. Their findings support the hypothesis that intense character motion and negative eye movement patterns are strong indicators of SS in VR environments.

While LLM-powered NPCs enhance user interaction through fluent dialogue and emotional tone, they often fall short in delivering sustained immersion due to limited animation, repetitive responses, and poor narrative memory. Studies show that visual feedback, avatar realism, and consistent personality cues are essential for believable interactions. However, fragmented storytelling and generic conversations remain common, especially over time. Achieving deeper engagement will require integrating expressive multimodal feedback, persistent memory, and emotionally aware behavior into LLM-driven virtual characters.

\subsection{Scalability and Deployment}

While many LLM-based prototypes in VR show promising capabilities, scaling them for real-world, large-scale applications remains a significant challenge. Transitioning from controlled lab settings to multiplayer or persistent virtual environments requires more than model performance—it demands robust infrastructure, cost-effective deployment, and compatibility with consumer hardware. As VR games become more complex and interactive, these demands intensify.

One of the main bottlenecks is the high computational cost of real-time LLM inference, especially in multi-agent settings where several NPCs must perceive, reason, and respond simultaneously. Techniques like retrieval-augmented generation and modular prompting aim to reduce memory load and latency, but their effectiveness is limited in fast-paced, interactive environments \cite{maslych2025takeaways, delaTorre2024llmr}. Multi-module systems like LLMR, while offering improved scene understanding, often face execution delays due to orchestration overhead, including planning, debugging, and memory updates \cite{delaTorre2024llmr}.

Hybrid system designs offer one potential solution. For example, combining LLMs with Pursuit Learning Automata (PLA) allows agents to learn user tone preferences and select pre-generated responses instead of generating them from scratch. This reduces processing load and supports smoother, long-term dialogue \cite{waghale2024ai, jahangiri2024balancing}. However, these methods are still in early stages and have only been tested in limited, single-user RPG setups, raising questions about scalability to multiplayer environments.

From a deployment standpoint, technical hurdles also persist. Many LLM-based VR systems rely on cloud APIs, leading to latency, privacy concerns, and service interruptions—all of which affect real-time performance. Integrating LLMs into engines like Unity often requires third-party middleware and custom tools, increasing development time and limiting cross-platform compatibility. While frameworks like SceneCraft showcase the narrative potential of LLMs, they still need optimization for real-time, in-engine deployment \cite{kumaran2023scenecraft}.

Memory and model management are also critical. Persistent agents must track long-term context, manage session memory, and coordinate with other agents. Current memory systems, such as those proposed by Buongiorno et al. and Zheng et al. \cite{buongiorno2024pangea, zheng2024memoryrepository}, improve dialogue continuity and increase system resource demands. These systems become unsustainable in larger, ongoing game worlds without efficient memory pruning, modular retrieval, and compression strategies.

Ultimately, scaling LLMs for real-time VR applications will require an integrated approach combining efficient local inference, modular design, optimized memory, and seamless engine integration. LLM-powered VR games will unlikely move beyond experimental prototypes into mainstream, persistent virtual worlds without these developments.

While LLMs demonstrate strong potential in VR game prototypes, their deployment at scale faces major challenges. High computational demands, latency from cloud APIs, and orchestration overhead limit real-time performance, especially in multi-agent or multiplayer settings. Hybrid systems and modular memory frameworks offer partial solutions, but they remain largely untested in large-scale environments. To transition from experimental setups to persistent virtual worlds, future systems must integrate lightweight inference, robust memory management, and native engine compatibility for seamless, scalable deployment. Beyond these LLM-specific challenges, users are also confronted with cybersickness—a more general limitation of VR environments, and this physiological discomfort can severely impact immersion.

\section{Future Directions}

As the integration of LLMs into VR games continues to evolve, several key areas emerge that will shape the next generation of immersive, intelligent, and ethically responsible digital experiences. This section outlines promising future directions based on current trends and gaps identified throughout this review.

\subsection{Advances in AI for More Realistic VR Gaming}

One of the most promising developments lies in the evolution of multimodal AI models that integrate language, vision, and audio understanding to support holistic, context-aware interaction. Systems such as GPT-4V \cite{openai2024gpt4} and LLaVA \cite{liu2023visualinstructiontuning} already demonstrate the capacity to interpret both images and text, which, when integrated into VR environments, can lead to NPCs that "see" and "hear" alongside players. These systems could track user gaze, interpret gestures, analyze visual scenes, and generate emotionally resonant, non-verbal responses in real time. Future advancements may enable avatars capable of full-body interaction understanding, dynamic emotion simulation, and persistent spatial memory across scenes, leading to virtual characters that feel more genuinely human and socially intelligent.

Furthermore, integrating LLMs with emotion recognition and affective computing will likely enhance their capacity for emotionally attuned interactions \cite{marin2018affective, harris2023eyetracking}. By incorporating data from facial expression tracking, voice tone analysis, and physiological inputs (e.g., heart rate, EEG), NPCs could dynamically respond to player moods, stress levels, or engagement patterns, enabling deeper player-character connections, particularly in therapeutic or educational applications \cite{pei2024eeg}.

\subsection{Integration with Other Technologies}

Beyond improved realism, future systems will benefit from the convergence of LLMs with complementary AI technologies. For instance, reinforcement Learning (RL) can be used to refine NPC behavior through iterative experience-based optimization, allowing characters to adapt over time and learn from player interactions. With LLMs’ generative flexibility, RL could support characters who evolve with players’ styles, forming long-term bonds or gameplay strategies \cite{justesen2019deep}.

Procedural content generation (PCG) is another area where LLMs can synergize with rule-based or stochastic systems to produce personalized worlds, quests, and story arcs in real time. Instead of replacing traditional PCG algorithms, LLMs can enrich them by filling narrative, dialogue, and context-sensitive decision trees with fluid, naturalistic language.

The Internet of Things (IoT) and wearable technology also offer integration potential for health-focused or location-aware VR experiences \cite{lv2023digital, marin2018affective}. For example, LLM-driven virtual coaches could adapt content or difficulty in real-time based on a player’s heart rate, body temperature, or environment, enhancing fitness, therapy, or safety applications. In a fitness game, if a player’s heart rate exceeds a safe threshold, an LLM assistant could reduce the workout intensity while explaining the reasoning and suggesting hydration tips.

\subsection{Ethical AI Development for Games}

With increasing immersion and personalization come rising ethical stakes. Future systems must embed ethical frameworks directly into the design pipeline, prioritizing bias mitigation, content moderation, explainability, and informed consent \cite{vainiopekka2023explainable}. This is especially urgent as AI characters become more emotionally responsive and persuasive, particularly among vulnerable populations such as children or neurodivergent users.

It will be key to building transparent AI systems that can explain their decisions, avoid reinforcing stereotypes, and flag potentially harmful outputs. This will also involve developing new benchmarking tools and ethical evaluation protocols that assess emotional impact, fairness, and long-term influence in gameplay, criteria often neglected in traditional usability testing. For example, a narrative-based VR simulation for conflict resolution teaching would use AI characters who modify their communication style and cultural references according to player background and emotional state to provide respectful guidance that avoids cultural or socioeconomic bias \cite{giaretta2025security}.

User data privacy must also be foregrounded. With VR systems collecting detailed biometric, motion, and interaction data, LLM-powered applications must ensure secure handling, precise opt-in mechanisms, and compliance with emerging privacy regulations \cite{natgunanathan2016protection}. Local processing or edge computing may reduce data transfer risks while enabling more responsive real-time AI.

\subsection{Open-Source and Industry Trends}

The growing open-source ecosystem is accelerating the democratization of LLM development and deployment. Models like LLaMA, Mistral, and Baichuan offer competitive performance and increased transparency, allowing smaller studios and researchers to experiment with AI-driven game mechanics without being locked into proprietary APIs. Open frameworks like LangChain, Hugging Face Transformers, and Unity GPT integrations also make it easier to develop modular, customizable LLM agents tailored to specific gameplay scenarios \cite{hu2025survey}.

In parallel, industry leaders like Meta, NVIDIA, and OpenAI actively invest in AI-native game engines and toolkits for generative NPCs, suggesting that large-scale, real-time LLM integration will soon become commercially viable. The release of real-time streaming APIs, quantized model deployment solutions, and multimodal interfaces will further reduce latency and memory constraints, facilitating scalable use in complex multi-agent virtual worlds.

Future research collaborations between academia and industry will be essential to ensure that these tools remain innovative, ethical, and inclusively designed. The open-source and commercial sectors should align around shared principles of transparency, accessibility, and responsible AI deployment.

\section{Conclusion}

This paper has examined the transformative role of Large Language Models in reshaping the landscape of Virtual Reality games. Through a comprehensive review of 62 recent studies, we analyzed the integration of LLMs across key domains: dynamic NPC interactions, procedural storytelling, intelligent game mastering, personalized experiences, accessibility design, and performance-aware deployment.

Our findings indicate that LLMs significantly expand the design space for VR games, enabling more adaptive, expressive, and emotionally resonant virtual characters. They support unscripted narrative branching, real-time user interaction, and socially intelligent behavior that redefines immersion. Moreover, LLMs open new pathways for inclusive and accessible VR systems, offering tailored experiences to diverse user populations, including those with disabilities or language barriers.

However, significant challenges remain. Real-time responsiveness, memory management, and latency limit large-scale deployment in complex VR scenarios. Ethical concerns—such as bias, privacy, and emotional manipulation—are amplified by the immersive nature of these experiences and require careful consideration throughout the design and testing processes. Scalability also presents a significant barrier as models must be optimized for resource-constrained hardware and sustained multi-agent interaction.

Future research should focus on advancing multimodal and hybrid AI systems, integrating reinforcement learning, affective computing, and spatial awareness. Developers and researchers must prioritize ethical AI development by embedding fairness, transparency, and safety into game mechanics and content generation pipelines. Collaboration across open-source communities and industry partners will be essential to make intelligent, inclusive, and creative VR experiences more accessible and impactful.
By combining technical innovation with responsible design, LLMs have the potential to fundamentally reshape how we build, experience, and interact within virtual worlds—ushering in a new era of intelligent, human-centered digital play.

\section*{Acknowledgment}

This research was financially supported by the TUM Campus Heilbronn Incentive Fund 2024 of the Technical University of Munich, TUM Campus Heilbronn. We gratefully acknowledge their support, which provided the essential resources and opportunities to conduct this study. 

\balance
\bibliographystyle{elsarticle-num}
\bibliography{LLM-paper}

\end{document}